\documentstyle[aps, preprint]{revtex}
\begin{document}
\preprint{MKPH-T-98-13}
\input{psfig.sty}
\title{ Anomalous Hypercharge Axial Current And The Couplings Of
The $\eta $ And f$_1$(1420) Mesons To The Nucleon}
\author{S. Neumeier$^1$ and M. Kirchbach$^2$\\
$^1${\small \it  Institut f\"ur Theoretische Physik, 
Universit\"at Leipzig, 
D-04109 Leipzig, Germany}\\
$^2${\small \it Institut f\"ur Kernphysik, J. Gutenberg Universit\"at, 
D-55099 Mainz, Germany} }
\maketitle
\begin{abstract}
It is emphasized that the precise three flavor symmetry of hadrons is
not SU(3)$_F$ but rather U(4)$_F$ restricted to
SU(2)$_{ud}\otimes $SU(2)$_{cs}\otimes $U(1) and considered in the limit
of frozen charm degree of freedom.
Within this scheme the hypercharge generator is
necessarily an element of the
su(2)$_{cs}\otimes $u(1) subalgebra as it contains the baryon number
generator associated with the unit matrix.
The structure of hypercharge obtained in this way is 
the only one that is consistent with the Gell-Mann--Nishijima relation.
In considering now the corresponding axial hypercharge transformations, 
the unit element of the u(4) algebra will give rise to the anomalous 
U(1)$_A$ current and the resulting hypercharge axial current will be 
anomalous, too. It is shown that the only anomaly free neutral
strong axial current having a well defined chiral limit is
identical (up to a constant factor) with the weak axial current.
There, a purely strange axial current comes in place of the hypercharge 
one and the $\eta $ meson acquires features of a `masked'
strange Goldstone boson. The consequence is that 
the $\eta N$ and f$_1$(1420)$N$ couplings have to proceed via
a purely strange isosinglet axial current. Therefore,
the $\eta $ and f$_1$(1420) mesons probe at the tree level the 
polarization of the strange quark sea and their couplings appear
strongly suppressed relative to quark model predictions.
{}For this reason, loop vertex corrections acquire importance. 
A model based on effective lagrangians for coupling the 
f$_1$(1420) and  $\eta $ mesons to the nucleon via triangular vertex 
corrections containing two--meson states
has been developed and shown to be convenient for data description
beyond the limits of applicability of chiral perturbation theory.


\noindent
KEY words: flavor symmetry, hypercharge axial current, U(1)$_A$ anomaly, 
           $\eta $ and f$_1$ meson-nucleon couplings  

\end{abstract}

\newpage

\section{Introduction}

Three flavor SU(3)$_F\otimes $SU(3)$_F$ chiral symmetry (ChS) of 
strong interaction is one of the basic guidelines in constructing the 
nucleon dynamics as it is considered to be the global internal symmetry of 
the QCD lagrangian \cite{Coleman}. As long as ChS is supposed to be realized 
in the non--multiplet Nambu--Goldstone mode with the pseudoscalar octet mesons 
acting as the associated Goldstone bosons, it becomes possible to expand 
correlation functions in powers of the current quark masses
and the external Goldstone boson momenta thought to be small at the
hadronic scale of $\Lambda \sim 1$ GeV \cite{GaLe}. This so--called
chiral perturbation theory (ChPT) advanced in the last decade to a 
powerful scheme for describing various low--energy phenomena \cite{Bern}.   
In view of this success, establishing the Goldstone boson character
of the lowest pseudoscalar mesons attracts special attention.

The first aim of the present study is to emphasize that the $\eta $ meson 
cannot be considered as the flavor octet Goldstone boson
but is rather a 'would be' strange Goldstone boson, a result
already conjectured in the previous work \cite{KiWe}.
The second is to suggest a phenomenological field theoretical
approach for describing meson--nucleon vertices beyond the limits of 
applicability of ChPT.
The reason for the non--octet Goldstone boson nature of
the $\eta $ meson is that according to an observation reported in
Ref.~\cite{KiPRD}, the precise three flavor symmetry of hadrons is
not SU(3)$_F$ but rather U(4)$_F$ restricted to
SU(2)$_{ud}\otimes $SU(2)$_{cs}\otimes $U(1) and considered in the limit
of frozen charm degree of freedom.
In such a case, the only anomaly free neutral strong axial current of
a well defined chiral limit is obtained in forbidding the U(1)$_A$
axial transformation. This current has same flavor 
structure as the neutral weak axial current.
Thus, within the three flavor space, a purely strange isosinglet axial 
current comes in place of the flavor octet axial current, and
no hypercharge Goldstone boson is required any more.
Rather, a purely strange Goldstone boson should be introduced, whose 
appearance is however prevented through the violation of the OZI rule for 
the pseudoscalar mesons as brought about by the U(1)$_A$ anomaly.

The consequence is that the neutral axial and pseudoscalar mesons will couple 
to the nucleon via the $\bar s s $ quarkonium components of their wave 
functions. The corresponding point-like vertices will be proportional to 
$\Delta s$, the small fraction of proton helicity carried by the strange quark 
sea, and appear strongly suppressed relative quark model predictions.
In view of that smallness, loop vertex corrections acquire importance. 

The effect of the one--loop corrections on the experimentally
observed strong suppression of the nucleon matrix element of the 
octet (or, hypercharge) axial vector current has been considered, for example, 
in Ref.~\cite{Sav} within the framework of SU(6) chiral perturbation theory. 
There, the strong dependence of the result on the $N$--$\Delta $ mass 
splitting was revealed and the necessity for higher-order corrections 
discussed. Instead of considering the suppression of the nucleon
octet axial matrix element, we here rather examine the
small enhancement of the axial $(\bar s s )N$ coupling in terms of
triangular corrections of the type  $a_0 (980) \pi N$ to the 
$\eta N N$-,  and  $K^*(892) KY$ to the f$_1$(1420)$NN$ vertices, 
respectively.  
Such triangular vertices participate
effective $Z(\bar s s) \, N $ chain couplings. Indeed, the coupling of an 
external weak isosinglet axial current to the nucleon can be viewed as being
mediated by the f$_1$(1420) or $\eta $ isosinglet meson states 
which contain strange and non-strange quarkonia simultaneously, thus violating 
the OZI rule.
In such a case, the $Z$ boson has at the weak vertex the opportunity
to select the strange $\bar s s $ quarkonium from the wave function of the 
respective intermediate meson, while at the strong vertex the nucleon 
can couple to the remaining $(\bar uu +\bar d d)$ quarkonia 
via its meson cloud as parametrized by the triangular vertices introduced 
above.

A method similar in spirit to the one presented here but of different
techniques is the so--called meson cloud model of Ref.~\cite{Szczurek}.
There, the authors study the influence of the mesons surrounding the
nucleon on its electromagnetic and weak vector form factors 
in terms of a nucleon wave function having a 
non--negligible overlap with the nucleon-meson scattering continuum.
We here focus rather on the isoscalar axial form factor. 

The paper is organized as follows. In the next section we consider 
the identification criteria for the Goldstone boson
character of the $\eta $ meson, discuss the ambiguities related to the 
calculation of the $\eta $ meson pole term, 
and show that the {\it axial hypercharge current is anomalous\/}. 
In Sect.\ 3 the contact $\eta NN$ and f$_1$(1420)$NN$ vertices
are considered. In Sect.\ 4 the effective $\eta NN$ vertex is calculated 
{}from the $\pi a_0(980)N$ triangular vertex, while Sect.\ 5 contains
the results on the effective f$_1$(1420) meson nucleon-couplings associated 
with the $KK^*(892) Y$ triangle. The paper ends with a short summary.

\section{$\eta $ Goldstone Boson Revisited}

The $\eta $ meson is canonically considered as the octet Goldstone
boson of three flavor SU(3)$_L\otimes $SU(3)$_R$ chiral symmetry as the 
sum of its pole term current, displayed in Fig.\ 1, and the hypercharge 
axial current of the nucleon ($N$) is assumed to be partially conserved 
\cite{deAlf}. The basic tacit assumption entering the calculation of 
the $\eta $ pole term is the universality of the hypercharge axial current 
at both the strong and weak vertices. Let us consider as an illustration 
the canonical ansatz according to which both the neutral flavor axial 
current and the $\eta $ meson wave function are approximately determined by 
Gell-Mann's hypercharge matrix $\lambda^8$, i.e. 
\begin{eqnarray}
|\eta \rangle \approx  \bar q_3\,  {\lambda^8\over \sqrt{2}} \, q_3\, ,
&\quad & j_{\mu ,5}^8 = \bar q_3 \, \gamma_\mu \gamma_5 \, 
{\lambda^8\over 2}\, q_3 \, ,\quad
\lambda^8 = {1\over \sqrt{3}}\left(\begin{array}{ccc}
1&0&0\\
0&1&0\\
0&0&-2\end{array}\right)\, . 
\label{P_meson}
\end{eqnarray}
We denote by $q_3$ the quark field in three flavor space: 
$q_3=(u\, d\, s)^T$.
The $\eta $ weak decay current is now parametrized as
\begin{equation}
J_{\mu ,5}^\eta \, := \langle 0|j_{\mu , 5}^8 |\eta \rangle = 
f_\eta \, iq_\mu \, .
\label{decay_const}
\end{equation}
Here, $q_\mu $ and $f_\eta $ stand for the four 
momentum and weak decay coupling constant of the $\eta $ meson, 
respectively.
In inserting Eq.~(\ref{P_meson}) into (\ref{decay_const}) and 
assuming in accordance to Ref.~\cite{Jaf89}
the quarkonia-quark currents couplings to be diagonal in flavor, i.e. 
\begin{equation}
\langle 0 |{1\over 2} \, \bar q q \, |\bar q' q' \rangle = 
\delta_{qq'} \kappa_q\, (0^-),
\quad q,q'=u,d,s \, ,
\label{kappa}
\end{equation}
the value of $f_\eta $ is calculated as 
\begin{equation}
3\sqrt{2} f_\eta=
\kappa_u (0^-) +\kappa_d (0^-) +4\kappa _s (0^-)\, .
\label{decay_constant}
\end{equation}     
Within this scheme, the contact octet meson--nucleon  vertex 
${\cal V}^8_{\eta N N }$ is given the form
\begin{eqnarray}
{\cal V}^8_{\eta N N  } =  {1\over f^2_\eta } \, 
                      J^{8 }_{5}\,\cdot  J^\eta_5
&=& {1\over f_\eta ^2 }
\left( G_A^{8}\, \bar{N}_2\gamma\gamma_5 
{\openone\over 2} N_1 \right) 
\,\cdot \, \left(f_\eta \, iq\,  \phi_\eta \right) \, .
\label{vert_PNN}
\end{eqnarray}
Here, $\phi_\eta $ stands for the field of the $\eta $ meson, 
while $N_1$ and $N_2$ denote the spinor fields of the in-- and outcoming
nucleons, respectively. The parameter of
the dimension $\lbrack mass^2 \rbrack $
entering the contact current--current coupling  
equals the weak $\eta  $ meson decay coupling constant. 
This choice implies the {\it current universality\/} mentioned above,
as it allows one to express the coupling of the $\eta $ meson to arbitrary 
baryon targets in terms of its weak decay coupling $f_\eta $.

The quantity $G_A^{8}$ in (\ref{vert_PNN}) stands for the coupling 
of an external axial lepton current to the nucleon hypercharge axial
current $J_{\mu ,5}^{8}$ and is defined via
\begin{equation}
s^\mu J_{\mu ,5}^{8}:= s^\mu \langle N_2| j_{\mu ,5}^8 |N_1\rangle 
= G_A^{8}\,
\bar {\cal U}_{N_2} \gamma_\mu \gamma_5  \, {\openone\over 2}
{\cal U}_{N_1}\,s^\mu  , \quad s^\mu q_\mu = 0\, ,
\label{G_A} 
\end{equation}
with $s^\mu $ being a unit spin polarization vector and ${\cal U}_N$
denoting nucleonic spinors.
The value of the nucleon hypercharge coupling $G_A^8$ following from
Eq.~(\ref{G_A}) is expressed in terms of the helicity fractions $\Delta u$,
$\Delta d$, and $\Delta s$, carried by the respective $u,d$, and $s$ quarks as
\begin{equation}
G_A^{8} = {1\over {\sqrt{3}} }(\Delta u+\Delta d 
- 2\Delta s)\, . \label{GA_h.fractions}
\label{GAheli}
\end{equation} 

The combination ${G_A^{8} \over {2f_\eta }}$ in 
Eq.~(\ref{vert_PNN}) is conventionally denoted
by ${f_{\eta NN }\over m_\eta } $ with $f_{\eta NN }$ being called the
gradient (or pseudovector (PV)) $\eta N$ coupling constant i.e.
\begin{eqnarray}
{f_{\eta NN } \over m_\eta } &=&{ G^{8}_A\over {2f_\eta  } }\, 
= { \sqrt{3\over 2}} {{\Delta u + \Delta d -2 \Delta s}\over
{\kappa_u (0^-) +\kappa_d (0^-) +4\kappa_s (0^-) }}\, ,
\label{currwf_mapp}
\end{eqnarray}
where use has been made of Eqs.~(\ref{decay_constant}) and 
(\ref{GA_h.fractions}).
On-mass shell, the PV coupling can be expressed through 
the pseudoscalar (PS) coupling $g_{\eta NN }$ via the equivalence
relation\cite{deAlf} as
\begin{eqnarray}
{f_{\eta NN }\over m_\eta } &=&{g_{\eta NN }\over {2m_N }}\, 
                   ={ G_A^{8}\over {2f_\eta }}\, \label{PV_PS}\, ,
\label{equiv_rel}
\end{eqnarray}
with $m_{N}$ standing for the nucleon mass.
The constraint on $g_{\eta NN }$ in Eq.~(\ref{PV_PS}) is known
as the hypercharge `Goldberger-Treiman' (GT) relation. 
Ordinarily, the GT constraint is alternatively obtained rather
as a condition sufficient for the partial conservation of the
hypercharge axial current already at the tree level \cite{deAlf} 
and is indicative for the Goldstone-boson character of 
the $\eta $ meson. The considerations given so far illustrate
that it is, actually, the assumed universality of the hypercharge
axial current at both the strong and weak vertices of the $\eta $ pole 
term which underlies the hypercharge GT-relation and thus enables the 
realization of hypercharge chiral symmetry in the hidden 
Nambu--Goldstone mode, and vice versa. 
In noting now that the weak neutral axial current 
$J_{\mu ,5}^w$, 
\begin{equation}
J_{\mu ,5}^w = -{1\over 2}
(\bar u \bar d)\gamma_\mu\gamma_5 {\tau_3\over 2}\left(\begin{array}{c}
u\\
d\end{array}\right) +{1\over 4}\bar s \gamma_\mu\gamma_5 s\, . 
\label{weak_current}
\end{equation}
apparently differs from $J_{\mu ,5}^8$,
one immediately realizes that  the universality assumption
brings an element of ambiguity in the calculation
of the $\eta $ pole term.
Indeed, in assuming the $\eta $ meson to couple to the hadronic vacuum via 
the hypercharge axial current (cf. Eq.~(\ref{decay_const})), 
one ascribes to this meson the 
ability to decompose the fundamental Z-boson current into hypercharge and 
flavor singlet components, much like an optically active material decomposes
a linearly polarized wave into its circularly polarized parts.
In that sense the neutral axial and pseudoscalar mesons are considered
as a sort of `electroweak active' probes.
This means that a constituent symmetry is given higher priority 
over a gauge symmetry without any deeper
justification. Remarkably, the assumed 
priority does not find any confirmation by data which speak in favor of a 
surprisingly small $\eta N$ coupling rather than in favor of the 
corresponding GT--prediction (see \cite{KiWe} for details)
even after accounting for the small $\eta -\eta '$ mixing required by 
the Gell-Mann-Okubo mass formulae. That such a mixing will lead to a reduction
of the nucleon matrix element of the octet axial current was earlier
considered, among others, in Ref.~\cite{Ven}. There, however,
the reduction reported was far away from being as big as the required one. 
Note, however, that the accuracy of GT--relations has been reliably proven
only for the case of the pion--nucleon system \cite{Coon}.
Below we argue that it is actually the electroweak axial 
current that  enters the calculation of pole terms containing
neutral pseudoscalar and axial vector mesons.
The argumentation given below essentially  follows Ref.~\cite{KiPRD}.

Consider the fundamental four flavor vector current of the quarks
\begin{eqnarray}
j_\mu & =&  {2\over 3}\bar u \gamma_\mu u 
               -{1\over 3} \bar d \gamma_\mu d
               +{2\over 3} \bar c \gamma_\mu c
               -{1\over 3} \bar s \gamma_\mu s\, \nonumber\\
&=&\bar q\,  t_3 \gamma_\mu q 
+\bar q {Y\over 2}\gamma_\mu q \, ,
\quad q=(u\, d\, c\, s)^T\, . 
\label{elm_curr}
\end{eqnarray}       
Here, $Y$ and $t_3$ in turn represent hypercharge and
third projection of isospin within the
quark flavor quadruplet, $q$, and are explicitly given below as
\begin{equation}
t_3 ={1\over 2} \left(\begin{array}{cccc}
1&0&0&0\\
0&-1&0&0\\
0&0&0&0\\
0&0&0&0
\end{array}\right)\, , \qquad 
Y=
\left(\begin{array}{cccc}
0&0&0&0\\
0&0&0&0\\
0&0&\hat{C}&0\\
0&0&0&\hat{S}\end{array}\right) + { {1\!\!1_4}\over 3} \, ,
\label{lambda_u4}
\end{equation}
with $\hat{C}c = c$, and $\hat{S}s=-s$, respectively.
As long as $j_{\mu }$ is conserved, its total charge
\begin{equation}
Q (t) = \int  j_0(t, \vec{x}\, )\mbox{d}^3\vec{x}\, ,
\label{octet_current}
\end{equation}
is a constant of motion and labels the hadron states.
When considered as an operator, $\hat{Q}$ is directly read off from
Eq.~(\ref{elm_curr})
to be related to the operators of isospin $\hat{t}_3 $
and hypercharge $\hat{Y}$ via the famous Gell-Mann--Nishijima relation 
\begin{eqnarray}
\hat{Q} &=& \hat{t}_3 +{1\over 2}\hat{Y}\, . 
\label{GLM_NSHI}
\end{eqnarray}
Now if Gell-Mann's matrix $\lambda^8$ in Eq.~(\ref{P_meson})
were to be interpreted as the generator of three flavor hypercharge,
it should emerge in the limit of negligible $c$ quark effects from
the complete four flavor hypercharge $Y$ of Eq.~(\ref{lambda_u4}) as
\begin{equation}
{1\over 3}\, 
\left(\begin{array}{cccc}
1&0&0&0\\
0&1&0&0\\
0&0&0&0\\
0&0&0&-2\end{array}\right)
=\lim_{m_c \to \Lambda_c} 
\left(\begin{array}{cccc}
0&0&0&0\\
0&0&0&0\\
0&0&\hat{C}&0\\
0&0&0&\hat{S}\end{array}\right)
+ {1\over 3}\, \lim_{m_c\to \Lambda_c }\, 1\!\!1_4\, .
\label{u4_lambda8}
\end{equation}
Here $\Lambda_c $ has to be sufficiently large in order to
`freeze out' the charm degree of freedom on the $1$ GeV mass scale, 
on the one side, and still finite, in order to preserve
the anomaly free character of the SU(4)$_F$ theory
\cite{Ambjoern}, on the other side. 
{}From the latter equation one sees that 
the matrix $\lambda^8$ representing the hypercharge in the 
truncated flavor space (now three dimensional) happens by accident to be
traceless while the full four flavor hypercharge matrix
$Y$ in Eq.~(\ref{lambda_u4}) is not traceless as it contains the unit matrix 
corresponding to the baryon number current. 
In this way the misleading impression appears that
hypercharge can be introduced on the level of the  group SU(3)$_F$
and be exploited for the construction of a partially conserved axial current. 
In contrast to Eq.~(\ref{u4_lambda8}), Gell-Mann's hypercharge matrix 
$\lambda^8$ from the su(3) algebra decomposes into 
\begin{eqnarray}
\sqrt{3}\lambda^8 &=& \lambda^3 +2\lambda^3_U\, ,\nonumber\\
\lambda^3 =\left(\begin{array}{ccc}
1&0&0\\
0&-1&0\\
0&0&0\end{array}\right)\, , &\quad &
\lambda^3_U =\left(\begin{array}{ccc}
0&0&0\\
0&1&0\\
0&0&-1\end{array}\right)\, ,
\label{Weyl_basis}
\end{eqnarray}
with ${1\over 2}\lambda^3 $ and ${1\over 2}\lambda^3_U  $ in turn denoting 
third projections of isospin and $U$-spin%
\footnote{The matrices $\lambda^3$ and $\lambda^3_U$ are nothing but
the diagonal elements of the su(3) algebra in the so-called Weyl basis.}. 
Looking at Eq.~(\ref{Weyl_basis}), one finds the following
expression for the electric charge of the quarks,
\begin{equation}
Q = t_3  +{1\over 3}(t_3 +2t_3^U)\, ,
\label{not_hypercharge}
\end{equation}
with
\begin{equation}
t_3^U ={1\over 2} \left(\begin{array}{cccc}
0&0&0&0\\
0&1&0&0\\
0&0&0&0\\
0&0&0&-1
\end{array}\right)\, .
\end{equation}
Eq.~(\ref{not_hypercharge}) differs from the standard 
relation, $Q=t_3 + (S+B)/2$, in group theoretical aspects.
It may be recalled that the 
full Gell-Mann--Nishijima relation, $Q=t_3 + (S+C+B)/2$ is a genuine 
u(1)$_B\oplus$su(4)$_F$-relation with the sum $S+C+B$ being 
the four flavor hypercharge $Y$. In the limit of neglected 
$c$ quark effects, this relation reduces to the form 
$Q=t_3 + (S+B)/2$ which, however, 
cannot be read as an su(3)$_F$-relation since the baryon number $B$ becomes
extraneous to this context. Thus, $(S+B)/2$ does not correspond to any 
su(3)$_F$ generator,
none of the su(3)$_F$ algebra 
elements can be given the interpretation of a hypercharge generator 
in the limit case of three flavors.
The physical hypercharge generator 
appears as a non--traceless element of the vector space
spanned by the u(4)$_F$ algebra and the corresponding
hypercharge axial current is anomalous. 
Indeed, in considering axial 
hypercharge transformations, the term containing the unit matrix on the rhs 
of Eq.~(\ref{u4_lambda8}) will give rise to the anomalously divergent 
U(1)$_A$ current for which no chiral limit can be formulated \cite{anom}. 
As a consequence, the hypercharge axial current will be {\it anomalous\/},
too. The only neutral flavor axial current which appears to be conserved in
the chiral limit of vanishing quark masses will be $j_{\mu ,5}$ defined as 
\begin{equation}
j_{\mu ,5} =   \bar q\,  \gamma_\mu\gamma_5 
\, \left( t_3+ {1\over 2}(Y -{{1\!\!1_4}\over 3}) \,  \right) \, q\, .
\label{weakf_curr}
\end{equation}
The flavor structure of $j_{\mu , 5}$ in the last equation  
reflects the exclusion of the anomalous U(1)$_A$ current \cite{anom} 
which cannot be used any longer as a building block for the construction 
of an anomaly free octet axial current. One remarkable feature of 
$j_{\mu ,5}$ is that its structure 
is identical (up to the factor of $-1/2$) to that of the neutral weak 
axial vector current and respects the OZI rule. 
{}For this reason, the well established universality of the flavor changing
weak and strong axial vector currents underlying the current algebra can be 
extended to include the neutral ones.
It is this current which will enter the calculations of
pole terms created by neutral pseudoscalar and axial vector mesons,
and the above mentioned ambiguity is resolved by now.
The  $j_{\mu ,5}$  current decomposes in three flavor space
into an isovector ($j_{\mu , 5}^I$) 
and a purely strange SU(2)$_I$ isosinglet ($j_{\mu , 5}^s $) component 
\begin{eqnarray}
j_{\mu ,5} &= & j_{\mu ,5}^I + j^s _{\mu ,5}\, , \nonumber\\
j_{\mu ,5}^I = \bar q_3 {\lambda^3\over 2}\gamma_\mu q_3\, , &\quad &
j^s_{\mu ,5} = -\bar s \gamma_\mu\gamma_5 {1\over 2} s \, , 
\label{isov_isosc}
\end{eqnarray}
with $\lambda^3$ being the isospin Gell-Mann matrix 
$\lambda^3$= diag(1,-1,0).

A remark is worthy on the flavor structure of the $\eta $ meson
which is the pseudoscalar analogue to the vector meson $\phi $.
While $\phi $ is an almost pure strange quarkonium,
the wave function of the physical $\eta $ meson derived from fitting
the meson mass spectrum has a significant non-strange quarkonium
component according to
\begin{equation}
|\eta \rangle = \cos\, \epsilon\, (-|\bar s s\rangle)
-\sin\,  \epsilon\, {1\over \sqrt{2}}(|\bar u u +\bar d d\rangle)\, ,
\quad
\epsilon = - 45.4^\circ \, .
\label{eta_Weyl}
\end{equation}
Here, ${1\over \sqrt{2}}(\bar u u +\bar d d)$ is the singlet
U(2)$_I$ state. For $\epsilon =-\arctan 1/\sqrt{2}$ the octet scalar state
$|\eta_8\rangle = q_3{\lambda^8\over \sqrt{2}}q_3$ is reproduced.
Eq.~(\ref{eta_Weyl}) clearly illustrates that the wave function of
the $\eta $ meson transforms in accordance with a representation
of $\lim_{m_c\to \Lambda_c} $U(1)$_I\otimes $SU(2)$_{cs}$ rather
than as a genuine SU(3)$_F$ state. From this point of view the
predominantly scalar octet nature of the $\eta $ meson appears
as an artefact of the violation of the OZI rule for the pseudoscalar
mesons as brought about by the U(1)$_A$ anomaly \cite{Shuryak}
rather than through a fundamental underlying SU(3)$_F$
symmetry.

Correspondingly, the $\eta '$ function reads,
\begin{equation}
|\eta ' \rangle = \sin \epsilon\, (-|\bar s s\rangle)
+\cos\,  \epsilon\, {1\over \sqrt{2}}(|\bar u u +\bar d d\rangle)\, ,
\quad
\epsilon = - 45.4^\circ\, .
\label{etaprime_Weyl}
\end{equation}
The large value of the angle $\epsilon $ in the last two equations 
signals a much stronger violation of the OZI rule within
the pseudoscalar nonet as compared to the vector meson nonet,
where $\epsilon \approx 5^\circ$.

\section{ Contact  $\eta NN$ And f$_1$(1420)$NN$  Vertices}
\label{sec-Contact}
The absence of a hypercharge component in the anomaly free
neutral strong axial current does not contradict the fact 
that the structure of the $\eta $ meson as deduced from 
data fits by means of the Gell-Mann-Okubo mass formulae, 
deviates from the purely strange quarkonium.
It only has essential impact on  the flavor structure of the
vertices including neutral octet axial and pseudoscalar mesons such as the 
$\eta $ and f$_1$(1420) mesons to the nucleon. 
In other words, while the leading components of the
$\eta $ and f$_1$(1420) meson wave functions can still be the scalar 
and singlet octet states, respectively, the $\eta N N$- and f$_1$(1420)$NN$ 
vertices will rather be purely strange isosinglets than 
of $F$-type because both these mesons can couple only to the 
anomaly free strange isosinglet nucleon current given below as
\begin{equation}
J_{\mu ,5}^{s (N)} = \langle N| {1\over 2}\bar s \gamma_\mu\gamma_5 s
|N\rangle = G_1^s \bar {\cal U}_N\gamma_\mu\gamma_5{\openone\over 2} 
{\cal U}_N\, ,
\qquad G_1^s = \Delta s\, ,
\label{G1_s}
\end{equation}
which can happen only via their strange quarkonium ingredients.

In approximating, for simplicity, the flavor part of the wave functions 
of both the $\eta $ and f$_1$(1420) mesons by 
Eq.~(\ref{eta_Weyl}) without caring at the moment about the
concrete numerical value of the corresponding mixing angle,  their new
axial currents are now defined by
\begin{eqnarray}
J_{\mu , 5}^{s\,  (\eta )} =\langle 0|
         -{1\over 2} \bar s \gamma_\mu\gamma_5 s 
| \eta \rangle 
&=& f_\eta\,  i q_\mu\, , \quad f_\eta :=
 \cos \epsilon\, \kappa_s (0^-)\,  m_\eta  \, ,\nonumber\\
J_{\mu , 5}^{s\,  (f_1 )} =\langle 0|
         -{1\over 2} \bar s \gamma_\mu\gamma_5 s 
| f_1 \rangle 
&=& f_{f_1}\,  m_{f_1}^2\epsilon_{f_1,}\, _{\mu}\, , \quad
f_{f_1} := \cos\epsilon\, \kappa_s (1^+)\,   \, .
\label{eta_axcurr}
\end{eqnarray}
Here $\kappa_s (0^-)$ and $\kappa _s(1^+)$ denote
the {\it dimensionless\/} couplings of the 
strange quarkonium components of the pseudoscalar and axial vector
mesons to the strange axial vector current, respectively.
The quantities $f_{f_1}$, $m_{f_1}$, and $\epsilon_{f_1,}\, _\mu $ stand for
the weak decay coupling, the mass and the polarization vector of the 
f$_1$(1420) meson.
{}Furthermore, the expressions in Eq.~(\ref{eta_axcurr}) have been obtained
in assuming universal couplings of the $\bar u u$ and $\bar d d $
quarkonia to the quark axial currents,
i.e. $\kappa_u (0^-) = \kappa_d (0^-)$, and
$\kappa_u (1^+)=\kappa_d (1^+)$, to recover isospin symmetry
already at the tree level\footnote{ The empirically 
observed closeness of the $\pi $ and $\eta $ weak decay constants,
($f_\eta \approx 1.1 f_\pi $) does not
necessarily imply $\kappa_u (0^-) \approx \kappa_s (0^-)$ at the tree 
level.}. 
In accordance with Eq.~(\ref{vert_PNN}) the purely strange
isosinglet contact  $\eta NN $
and f$_1$(1420)$NN$ vertices now read 
\begin{eqnarray}
{\cal V}^s_{\eta N N } =  {1\over f^2_\eta } \, 
                       J^{s\, (N)}\cdot J^{s\, (\eta )}\,  
&=& {1\over f_\eta^2 } G_1^s\bar{N}\gamma\gamma_5 
{\openone\over 2} N \,\cdot \, f_\eta \, iq \phi_\eta\, ,\nonumber\\
{\cal V}^s_{f_1 N N } =  {1\over m_{f_1}^2 } \, 
                       J^{s\, (N)}\cdot J^{s\, (f_1 )}\, 
&=& {1\over m_{f_1}^2 } G_1^s\bar{N}\gamma\gamma_5 
{\openone\over 2} N \,\cdot \, f_\eta \,m_{f_1}^2\, \phi_{f_1}\, ,
\label{vert_etaNN}
\end{eqnarray}
with $\phi_{f_1}$ standing for the f$_1$ meson field.
The combination ${{G_1^sf_{f_1}}\over 2}$ is ordinarily identified with 
the f$_1$(1420)$N$ contact coupling $f_{f_1NN}$. 
In inserting Eqs.~(\ref{G1_s}) and (\ref{eta_axcurr}) in the last 
expressions, one is led to the following relations:
\begin{eqnarray}
{f_{\eta NN} \over m_\eta } ={ G_1^s\over {2f_\eta } }\, , \quad
&\mbox{with}& \quad 
f_{\eta NN}  = {{\Delta s}\over {2\cos\epsilon\, \kappa_s (0^-)}}\, ,
\nonumber\\
f_{f_1 NN}  ={ {G_1^s f_1}\over 2 }\, , \quad
&\mbox{with} & \quad 
f_{f_1 NN}  = {{\Delta s}\over 2}\,\cos\epsilon\, \kappa_s (1^+)\, .
\label{etaN_vert}
\end{eqnarray}     
In contrast to Eq.~(\ref{GA_h.fractions}), the polarization
of the non--strange sea does not any longer participate the weak decay
couplings of the $\eta $ and f$_1$(1420) mesons.
In exploiting the on--shell equivalence relation between pseudoscalar and 
pseudovector couplings \cite{deAlf} one finds for the pseudoscalar $\eta N$ 
coupling constant $g_{\eta NN}$ the following expression
\begin{equation}
g_{\eta NN} =  { f_{\eta NN} \over m_\eta  }\, 2m_N\, 
= {{\Delta s}\over {2\cos\epsilon\, \kappa_s (0^-)}}\, 
{{2m_N}\over m_\eta }\, .
\label{PS_etaN}
\end{equation}  
As long as Eq.~(\ref{etaN_vert}) links the $\eta N$ coupling to $G_1^s$
by means of a Goldberger-Treiman relation, the $\eta $ meson can be considered
as a `would be' strange Goldstone boson.  
This means that at tree level the contact $\eta N$ coupling 
appears proportional to the fraction of proton helicity carried by the 
strange quark sea, rather than to the octet axial vector coupling 
$G_A^{(8)} ={1\over \sqrt{3}}(\Delta u + \Delta d - 2\Delta s)$, 
a result already conjectured in a previous work \cite{KiWe}. 
Due to the smallness of $\Delta s =-0.08\pm 0.05  $ 
(see \cite{Ell93} for a recent review), the $\eta$ 
and f$_1$(1420) mesons will almost decouple at tree level from the nucleon.
This might be one of the main reasons for which a strong suppression of the 
$\eta N$ couplings has frequently been found over the years by various 
data analyses of $\eta $ photoproduction off proton at threshold 
\cite{TiKa}, $\bar p p $ collisions \cite{Kroll}, as well as 
nucleon-nucleon (NN) and nucleon--hyperon (NY) phase shifts \cite{Reuber}. 

One of the main points of the present study is that 
the tree level $\eta N$ and $f_1 N$ couplings almost vanish as they
proceed over a purely strange isosinglet axial current.
In the following, the small but non--negligible couplings of these
mesons to the nucleon will be entirely attributed to
triangular vertices of the type $a_0 \pi N$, and $KK^*Y$, respectively.

\section{Effective $\eta NN$ Vertices}
\label{sec-etaNN}
In this section we calculate the gradient and pseudoscalar 
coupling constants of the $\eta $ meson to the nucleon by means of triangular
vertices involving the $a_0$(980) and $\pi $ mesons.

The special role of the $a_0(980)\pi N$ triangular diagram
as the dominant one--loop mechanism for the $\eta N$ coupling 
is singled out by the circumstance that the $a_0(980)$ meson is
the lightest meson with a two particle decay channel
containing the $\eta $ particle \cite{PDG94}.
The contributions of heavier mesons such as
the isotriplet $a_2(1320)$ tensor meson with an $\eta\pi $ 
decay channel and the
isoscalar $f_0(1400)$, $f^\prime _2(1525)$ and $f_2(1720)$ tensor
mesons with $\eta \eta $ decay channels
will be left out of consideration
because of the short range character
of the corresponding triangle diagrams on the one side,%
\footnote{The same argument applies to
the neglect of the $f_0(1590)\eta N$ triangular
vertex.}
and because of the comparatively small couplings of the tensor
mesons to the nucleon \cite{Ma89,Els87} on the other side.

The $\pi a_0 N $ triangular couplings 
(Fig.\ 2)
have been calculated using the following effective lagrangians
of common use:
\begin{eqnarray}
{\cal L}_{a_0\eta \pi }(x) & = & f_{a_0\eta \pi } 
{{m_{a_0}^2 -m_\eta^2}\over m_\pi}\phi_\eta^\dagger (x) 
\vec{\phi_\pi}(x)\cdot
\vec{\phi}_{a_0}(x) \\
{\cal L}_{\pi NN}(x) & = & 
\frac{f_{\pi NN}}{m_\pi}\bar N (x)\gamma_\mu \gamma_5
\vec \tau N (x)\cdot \partial^\mu\vec {\phi}_\pi (x) , \\ 
{\cal L}_{a_0NN}(x)& = & g_{a_0 NN}\,i \bar N (x)
\vec\tau N (x)\cdot\vec{\phi}_{a_0}(x) \, .
\label{a0pi_larg}
\end{eqnarray}
Here $f_{\pi NN}$ and $g_{a_0NN}$ 
in turn denote the pseudovector $\pi N$ 
and  the scalar  $a_0N$ coupling constants. We
adopt for $f_{\pi NN}$ the standard value $f_{\pi NN}^2/4\pi = 0.075$ 
and fit  $g_{a_0 NN}$ to data. 
The value of $f_{a_0\eta \pi }=0.44$ has been extracted 
{}from the experimental decay width \cite{PDG94} when ascribing the total $a_0$
width to the $a_0\to \eta + \pi$ decay channel.
The amplitude $T_{\eta N}(\pi +N\to a_0+N )$ entering the diagram 
in 
Fig.\ 2
can be 
parametrized in terms of the following complete set of invariants
\begin{equation} 
T_{\eta N}(\pi +N\to a_0 +N ) 
= \bar {\cal U}_N(\vec{p}\, ') \left( {{G_1(k^2)}\over m_\eta }\, 
k\!\!\! / \gamma_5 + G_2(k^2)\, \gamma_5 \right)\, {\cal U}_N(\vec{p}\, )\, , 
\label{etaN_ampl}
\end{equation}
where the invariant functions $G_1 (k^2)$ and $G_2(k^2)$ in turn correspond 
to pseudovector (PV) and pseudoscalar (PS) types of the $ \eta N$ coupling.  
Expressions for $G_1(k^2)$ and $G_2(k^2)$ can be found in evaluating the 
diagrams in 
Fig.\ 3
in accordance with
the standard Feynman rules.
We here systematically consider the incoming proton to be on its mass
shell and make use of the Dirac equation, so that
\begin{equation}
\gamma_5 p\!\!\! / \,\, {\cal U}_N (\vec{p}\, )= m_N
\gamma_5\, {\cal U}_N(\vec{p}\, )\, ,
\label{pin_on2}
\end{equation}
holds. On the contrary, the outgoing proton has been considered 
to be off its mass shell with
\begin{equation}
\gamma_5 p\!\!\!/\, '  {\cal U}_N(\vec{p}\, ) =
\gamma_5 (p\!\!\!/+k\!\!\!/){\cal U}_N(\vec{p}\, )
=m_N\gamma_5{\cal U}_N(\vec{p})+
 \gamma_5k\!\!\!/\, {\cal U}_N(\vec{p}\, )\, .
\label{pout_off2}
\end{equation}

The final result on 
$G_1(k^2)$ obtained in this way reads:
\begin{eqnarray}
G_1(k^2) & = &
C\, \int_0^1\int_0^1 dy dx x {{c_1 (x,y,k^2)}\over
  { {\cal Z} (m_N, m_\pi,m_{a_0}, x,y,k^2 )} }\, , \nonumber\\
c_1(x,y,k^2)& =  & -{1\over 2} x(1-y)m_Nm_\eta  \, , \nonumber\\
C &=& \frac{3}{8\pi^2}{{m_{a_0}^2-m_\eta^2}\over m_\pi^2}
 f_{\pi NN}f_{a_0\eta \pi}g_{a_0 NN} \, . 
\label{f_eta}
\end{eqnarray}
The corresponding expression for $G_2(k^2)$ reads
\begin{eqnarray}
G_2(k^2) & = &
C\, \int_0^1\int_0^1 dy dx x {{c_2(x,y, k^2)}\over
{{\cal Z}(m_N, m_\pi,m_{a_0}, x,y,k^2)}}\, ,\nonumber\\
c_2(x, y, k^2) & =& -x(1-y)m_N^2 \, .
\label{g_PS} 
\end{eqnarray}
The function ${\cal Z}(m_B,  m_1,m_2, x,y,k^2)$  appearing in the last
two expressions is defined as
\begin{eqnarray}
{\cal Z}(m_B, m_1,m_2, x,y, k^2) & =& 
m_N^2x^2 (1-y)^2 +  x^2yk^2 + m_1^2(1-x) + (m_2^2- k^2)xy\nonumber\\ 
&+&(m_B^2-m_N^2)x(1-y)\, .
\end{eqnarray}

The remarkable feature of the analytical expressions for the
pseudoscalar and pseudovector $\eta N$ couplings is that
they are given by {\it completely convergent\/} integrals and depend
only on the $a_0\to \pi +\eta $ decay constant and the
respective pion and $a_0$ meson-nucleon couplings.
The sources of uncertainty in the parametrization of the
effective $\eta NN $ vertex by means of
the triangular $a_0(980)\pi N$ diagram are
associated with the $a_0(980) N$ coupling constant and
the $\Gamma (\eta \pi)/ \Gamma^{\rm tot}_{a_0}$
branching ratio. For example, the coupling constant $g_{a_0NN}$ varies 
between $\approx 3.11$ and $\approx 10$ depending on the $NN$ potential 
model version \cite{Ma89,Els87}. Because of that we give below
the values for the gradient and pseudoscalar $\eta N$ couplings 
following from the $a_0\pi N$ triangular $\eta N$ vertex 
as a function of the $a_0N$ coupling constant:
\begin{eqnarray}
|G_1 (k^2 =m_\eta^2)|  = 0.06 \, g_{a_0NN}\, , &\quad & 
|G_2 (k^2 =m_\eta^2 )| = 0.22 \, g_{a_0NN}\, .
\end{eqnarray}
There are the quantities in Eqs.~(\ref{f_eta}) and (\ref{g_PS})
which we shall interpret as the {\it effective\/}
 pseu\-dovec\-tor and pseudoscalar
$\eta N$ coupling constants, respectively,
\begin{eqnarray}
f^{\rm eff}_{\eta NN} \, =\,  G_1(k^2)\, ,&\qquad &
g^{\rm eff} _{\eta NN} \, = \, G_2 (k^2)\, .
\label{PVPS_cpl}
\end{eqnarray}
Data analyses on $\eta $ photoproduction off proton near threshold suggest
for the pseudoscalar $\eta N$ coupling the small value of
$g_{\eta NN}^2 /4\pi \approx 0.4$.
The value of $g_{a_0 NN}$ that fits this number
corresponds to the maximal magnitude of $g^2_{a_0 NN}/4\pi \approx 6.79 $
reported in \cite{Ma89}.
{}From Eq.~(\ref{etaN_ampl}) one sees that the triangular $a_0\pi N$ 
correction to the $\eta NN $ vertex represents a mixture
\cite{Gross} of pseudovector and pseudoscalar types of $\eta N$ couplings.
This mixing is quite important for reproducing the form of the
differential cross section for $\eta $ photoproduction off proton
at threshold in 
Fig.\ 5.
Note that such a mixing cannot take place for Goldstone bosons 
because their point like gradient couplings to quarks are determined
in an unique way. On the contrary, in case of extended effective 
meson-nucleon vertices, such a mixing can take place by means of 
Eq.~(\ref{etaN_ampl}).
Data compatibility with the PV-PS mixing created by the $\pi a_0 N$ triangular 
correction to the $\eta NN$ vertex is a further hint on the non--octet Goldstone 
boson nature of the $\eta $ meson
\footnote{It should be noted that the analytical expressions for 
the $\eta N$ couplings 
in Eqs.~(\ref{g_PS}) and (\ref{f_eta}) differ from those 
obtained in \cite{KiTi} where the ambiguity in treating the
off--shellness of the outgoing proton was not kept minimal.
In the present calculation, the on--shell approximation 
$p\cdot k =-p'\cdot k = -m^2_\eta /2 $ was made
only in evaluating the denominators of the Feynman diagrams, whereas
in the nominators $p\!\!\!/ ' $ was consequently replaced by
$p\!\!\!/ ' =\, p\!\!\!/ + \, k\!\!\!/ $. In contrast to this, in \cite{KiTi}
the above on--shell approximation was applied to the nominators
too and, in addition, terms containing $p\!\!\!/ '$ have been occasionally 
interpreted as independent couplings. Through the improper treatment of the
off-shellness of the outgoing proton in \cite{KiTi}, 
logarithmically divergent integrals have been artificially invoked 
in $g_{\eta NN}$.}.

\section{Effective f$_1$(1420)$NN$ Vertices}
\label{sec-f1NN}
The internal structure of the axial vector meson 
f$_1$(1420) is still subject to some debates (see Note on f$_1$(1420)
in \cite{PDG94}). Within the constituent quark model this meson
is considered as the candidate for the axial meson ($\bar s s$)
state and therefore as the parity partner to $\phi $ 
{}from the vector meson nonet. 
The basic difference between the neutral $1^-$ and $1^+$
vector mesons is that while the physical $\omega $ and $\phi $
mesons are almost perfect non-strange and strange quarkonia, respectively,
their corresponding parity partners f$_1$(1285) and f$_1$(1420)
are not. For these axial vector mesons strange and non--strange 
quarkonia appear mixed up by the angle $\epsilon \approx  15^\circ$:
\begin{eqnarray}
|f_1(1285) \rangle &=& -\sin\epsilon\, (|\bar s s\rangle)
+\cos\,  \epsilon\, {1\over \sqrt{2}}(|\bar u u +\bar d d\rangle)\, ,
\label{D_Weyl} \\
|f_1(1420) \rangle &=& \cos\, \epsilon\, (|\bar s s\rangle)
+ \sin\,  \epsilon\, {1\over \sqrt{2}}(|\bar u u +\bar d d\rangle)\, ,
\quad
\epsilon \approx  15^\circ\,  .
\label{E_Weyl} 
\end{eqnarray}
Therefore within this scheme the violation of the OZI rule
for the neutral axial vector mesons appears quite different as
compared to the pseudoscalar mesons where the angle corresponding 
to $\epsilon $ was found to be $\epsilon \approx -45^\circ$.
 
In other words, while the OZI rule is respected by the internal
{}flavor structure of the vector mesons 
(there is an almost complete separation between the strange and 
non--strange quarkonia), it is violated for the axial ones. 
The latter effect parallels the situation 
within the $0^-$ octet and is interpreted as the consequence of 
the U(1)$_A$ anomaly, a subject discussed in \cite{Shuryak}. 
On the other side, the f$_1$(1420) meson seems 
alternatively to be equally well interpreted as a $K^*\bar K$ molecule 
\cite{Kmol}. The coupling of this meson to the nucleon is not 
experimentally well established so far. The only information about it
 can be obtained from fitting $NN$ phase shifts  by means of generalized 
boson exchange potentials of the type considered in Refs.\cite{Ma89,Els87}. 
There, one finds that the coupling $g_{f_1 NN}\approx 10$ to the nucleon
of an effective f$_1$ meson having same mass as the 
{}f$_1$(1285) meson is comparable to the $\omega N$ coupling. 
Below we demonstrate that couplings of that magnitude can be associated to a 
large amount with triangular diagrams of the type $KK^*Y$.

To calculate the $KK^*Y$ triangular coupling in 
(Fig.\ 4)
we use the following effective lagrangians:
\begin{eqnarray}
{\cal L}_{f_1KK^* }(x) & = & f_{f_1KK^* } 
{{m_{K^*}^2 -m_K^2}\over m_K} \, 
K^\dagger (x) K^* (x)^\mu f_1 (x)_\mu + h.c.\, , \label{f1KKvertex}\\
{\cal L}_{K Y N}(x) & = & 
g_{KNY} \, i\, \bar Y (x)\gamma_5 N (x) 
 K (x) +h.c. , \\ 
{\cal L}_{K^* YN}(x)& = & 
-g_{K^*YN}\bar N (x) (\gamma^\mu K^* (x)_\mu 
+{{\kappa_V}\over {m_N+m_Y}}
\sigma^{\mu\nu} \partial_\nu K^*(x)_\mu )Y (x)   +h.c.\,  
\label{KK*_lagr}
\end{eqnarray}
Here, $g_{KNY}$ stands for the pseudoscalar coupling of the
kaon to the nucleon-hyperon ($Y$) system, $g_{K^*NY}$ denotes
the vectorial $K^* NY$ coupling, $ g_{K^*NY}{{\kappa _V}\over {m_N+m_Y}} $ 
is the tensor coupling of the $K^*$ meson to the baryon, $K^*(x)_\mu $, 
and f$_1(x)_\mu $ in turn stand for the polarization vectors of the $K^*$
and f$_1$ mesons, $K(x)$ is the kaon field, while $N(x)$ and $Y(x)$ are in turn
the nucleon and hyperon fields.
A comment on the construction of the effective 
langrangian ${\cal L}_{f_1KK^*}$ is in place. 
Ogievetsky and Zupnik (OZ) \cite{OZ} constructed
a chirally invariant lagrangian containing no more than two
{}field derivatives to describe the dynamics of the 
$a_1(1260) \rho(770) \pi$ system. Exploiting the fact that these mesons
have the same external quantum numbers $J^{PC}$ as the 
respective f$_1$(1420), $K^{\ast},$ and $K$ mesons,
the OZ lagrangian might serve as a model for
the f$_1  K^* K$ system. However, the calculation shows up
UV-divergent integrals which are to be maintained
by additional cut--offs, i.e.\  by introducing form factors 
at both the $KYN$- and $K^*YN$-vertices \cite{SN}. 
This inconvenience makes the lagrangian choice according to 
\cite{OZ} less attractive than the present one given above
by Eq.~(\ref{f1KKvertex}). 

We adopt for the coupling constants the values implied by the J\"ulich 
potential \cite{Reuber}
\begin{eqnarray}
{{g^2_{p\Lambda K^+}}\over {4\pi}}
= (-0.952)^2\, , &\quad & {{g^2_{p\Lambda K^*}}\over {4\pi}}
= (-1.588)^2\, ,\quad \kappa_V = 4.5\, .
\label{KK*_couplings}
\end{eqnarray}
The value for $f_{f_1KK^*}= 1.97$ has been extracted from the experimental
$\Gamma_{\bar K K^*+h.c.}$ width of 17 MeV \cite{PDG94}.
The amplitude $T_{f_1N}(K+N\to K^*+N) $ entering the
effective f$_1NN$ vertex can now be expanded into the complete
set of the following invariants 
\begin{eqnarray}
T_{f_1N} (K+N\to K^*+N) & = & \bar {\cal U}_N(\vec{p}\, ')\,  
(F_1 (k^2)\, \gamma\cdot \epsilon_{f_1}\gamma_5 +
{{F_2 (k^2)}\over m_{f_1}} \, 
\gamma \cdot \epsilon_{f_1}\gamma_5 \gamma\cdot k\, 
\nonumber\\
&+& {{F_3(k^2)}\over {m_Nm_{f_1}}}\,  
p\cdot \epsilon_{f_1}\gamma_5\gamma\cdot k 
+ {{F_4(k^2)}\over m_N}\, p\cdot \epsilon_{f_1} \gamma_5\, ) 
\, {\cal U}_N(\vec{p}\, )\, .
\label{f1_coupl}
\end{eqnarray}
It will become clear in due course that 
the vector part of the $K^*NY $ coupling contributes only to the 
$F_2(k^2)$ and $F_4(k^2)$ currents in Eq.~(\ref{f1_coupl}). All the remaining
terms are entirely due to the $K^*NY$ tensor coupling.
{}For small external momenta, i.e. when 
$p\!\!\!/'{\cal U}_N (\vec{p}\, ) \approx m_N{\cal U}_N(\vec{p}\, )$,   
the number of the invariants reduces to three as the 
{}first term in Eq.~(\ref{f1_coupl}) can be approximated by the 
{}following linear combinations of the $F_2(k^2)$ and $F_3(k^2)$ currents:
\begin{eqnarray}
m_N \bar {\cal U}_N (\vec{p}\,')\gamma_5 k\!\!\!/\gamma_\mu 
{\cal U}_N (\vec{p}\, )
= - p'\cdot k\, \bar {\cal U}_N(\vec{p}\, ) \gamma_5\gamma_\mu 
{\cal U}_N(\vec{p}\, )
&+& p_\mu ' \, \bar {\cal U}_N(\vec{p}\, )\gamma_5\, k\!\!\!/
\, {\cal U}_N (\vec{p}\, )\, .
\label{trick}
\end{eqnarray}
One can make use of the approximation of Eq.~(\ref{trick}) 
to estimate $F_1(k^2)$, the only invariant function which cannot be 
deduced in an unique way from the triangular $KK^*Y$ diagrams 
because of the (logarithmically) divergent parts contained there.

The invariant functions $F_i(k^2)$ are now evaluated by means of 
the diagrams of momentum--flow in 
Fig.\ 3
in using the techniques of the previous section.
The various f$_1 N$ couplings can then be expressed in terms
of the integrals
\begin{equation}
 F_i(k^2) = \frac{1}{16\pi^2} f_{f_1KK^*} g_{KYN}g_{K^*YN}\, \int_0^1\, 
x\mbox{d}x\int_0^1\mbox{d}y 
B_i(k^2)\, .
\label{f1_FF}
\end{equation}
The only divergent analytical expression is the one for
$B_1(k^2)$ 
\begin{eqnarray}
B_1(k^2) &=& 2\bar\kappa {{m_{f_1}^2-m_{K^*}^2}\over m_K}
m_N (xm_Y - 2m_N){1\over Z}\nonumber\\
&+& \bar \kappa k^2  {{m_{f_1}^2-m_{K^*}^2}\over m_K}
(xy+(1-x)(8xy+x-6)){1\over Z}
\, ,\nonumber\\
&-&2\bar\kappa {{m_{f_1}^2-m_{K^*}^2}\over m_K}
\ln { {{\cal Z}(m_Y,m_K,\Lambda_{K^*},x,y,k^2)
      {\cal Z}(m_Y,\Lambda_K,m_{K^*},x,y,k^2)}\over
     {{\cal Z}(m_Y,m_K,m_{K^*},x,y,k^2)
      {\cal Z}(m_Y,\Lambda_K,\Lambda_{K^*},x,y,k^2)} }\, .
\label{diverg_ff}
\end{eqnarray}
All the remaining invariants are convergent and given below as
\begin{eqnarray}
B_2(k^2) &=& 
{{m_{f_1}^2-m_{K^*}^2}\over m_K}
m_{f_1}[\bar\kappa (m_Y(-x^2y(1-y)+3xy-3x+2) + xym_N)+(1-x(1+y))]
{1\over Z}
\, , \nonumber\\
B_3(k^2)&=& -\bar \kappa m_{f_1}m_N{{m_{f_1}^2-m_{K^*}^2}\over m_K}
((1-y)(x^2(1+3y)-3x)+2)\, {1\over Z}
\, ,\nonumber\\
B_4(k^2) &=& -2m_N{{m_{f_1}^2-m_{K^*}^2}\over m_K}
[(x(1-y)-1)(1+\bar\kappa m_N x(1-y)) + \bar\kappa m_Y x(1+y) ] \,
{1\over Z}\, , \nonumber\\
Z&=& {\cal Z}(m_Y,m_K,m_{K^*},x,y,k^2) \, .
\label{B_is}
\end{eqnarray}
The numerical evaluation of the expressions in the last two equations
leads to the following results:
\begin{eqnarray}
F_1 (k^2=m_{f_1}^2) = -8.56 \, , &\quad &
F_ 2 (k^2=m_{f_1}^2) = 6.83\, ,\nonumber\\
F_3(k^2=m_{f_1}^2) = -3.77\, , &\quad &
F_4(k^2=m_{f_1}^2) = -2.03\, ,\nonumber\\
m_{f_1} &=& 1385.7\, {\rm MeV}. 
\label{Werte_Bis}
\end{eqnarray}
These results show that all the couplings in Eq.~(\ref{f1_coupl})
are significant and have to be taken into account in calculating
$\eta $ and f$_1$ meson production cross sections.
\section{Summary}
As soon as we take three flavor symmetry to emerge from U(4)$_F$
restricted to SU(2)$_{ud}\otimes $SU(2)$_{cs}\otimes $U(1)
in the limit of `frozen'  charm degree of freedom,
we showed that the hypercharge axial current is anomalous,
that the $\eta $ meson acts as a 'would be' strange Goldstone boson,
and that the contact $\eta N$ and f$_1 N$ couplings
proceed via a purely strange isosinglet axial current.
For this reason, the couplings considered appeared 
proportional to $\Delta s$, the small fraction of nucleon helicity carried by 
the strange quark sea and strongly suppressed relative quark model predictions.
Within this scenario loop vertex corrections acquired importance.
We used effective lagrangians to construct effective 
$\eta N$ and f$_1 N$ coupling strengths in terms of
triangular $a_0 \pi N$, and $KK^*Y$ diagrams, respectively.
We found all kinds of effective couplings (up to one) to be 
determined by divergenceless expressions. This is the reason
{}for which we view the model developed here as useful
{}for describing $\eta N$ and f$_1$(1420)$N$ couplings beyond the limits
of applicability of ChPT. The strengths of all the invariant amplitudes
contributing to the spatially extended triangular meson--nucleon vertices
were found to be of significant size
and the produced mixing between the different types of meson--nucleon
couplings was shown to be important for data interpretation. 
Especially the form of the differential cross section for $\eta $ 
photoproduction off proton at threshold was successfully reproduced 
in terms of 
a mixing between pseudoscalar and 
pseudovector $\eta N$ couplings brought about 
by the invariant decomposition of the $a_0 +N \to \pi + N$ scattering amplitude
entering the triangular $a_0\pi N$ vertex.

\vspace{0.1cm}
\noindent
{\bf Acknowledgements}

We wish to thank Hans-Joachim Weber for continuous supportive discussions
on the nature of the $\eta N$ coupling mechanism, Martin Reuter for
his interest and helpful remarks on particular group theoretical aspects,
and Andreas Wirzba for his comments on the 
U(1)$_A$ anomaly problems.

This work was partly supported by the Deutsche Forschungsgemeinschaft
(SFB 201).

\vspace{1cm}
\noindent
{\bf Figures} 
\\
\noindent

Fig.\ 1: \quad The $\eta $ pole term. 

\noindent

Fig.\ 2: \quad The triangular $\eta NN$ vertex. 

\noindent

Fig.\ 3: \quad The flow of momentum.  

\noindent

Fig.\ 4: \quad The triangular $f_1 NN$ vertex. 

\noindent

Fig.\ 5: \quad Differential cross section for $\eta $ photoproduction
off proton at lab energy of 724 MeV as calculated
within the model of Tiator, Bennhold and Kamalov ~\cite{TiKa}.
The dotted and dash--dotted lines correspond to 
$g_{\eta NN}^2/4\pi $ taking the values of 0.4 and 1.1, respectively.
 The full line corresponds to the $a_0\pi N$ triangular coupling.
The data are taken from Krusche et al. \cite{TiKa}.

\begin{figure}[htbp]
\centerline{\psfig{figure=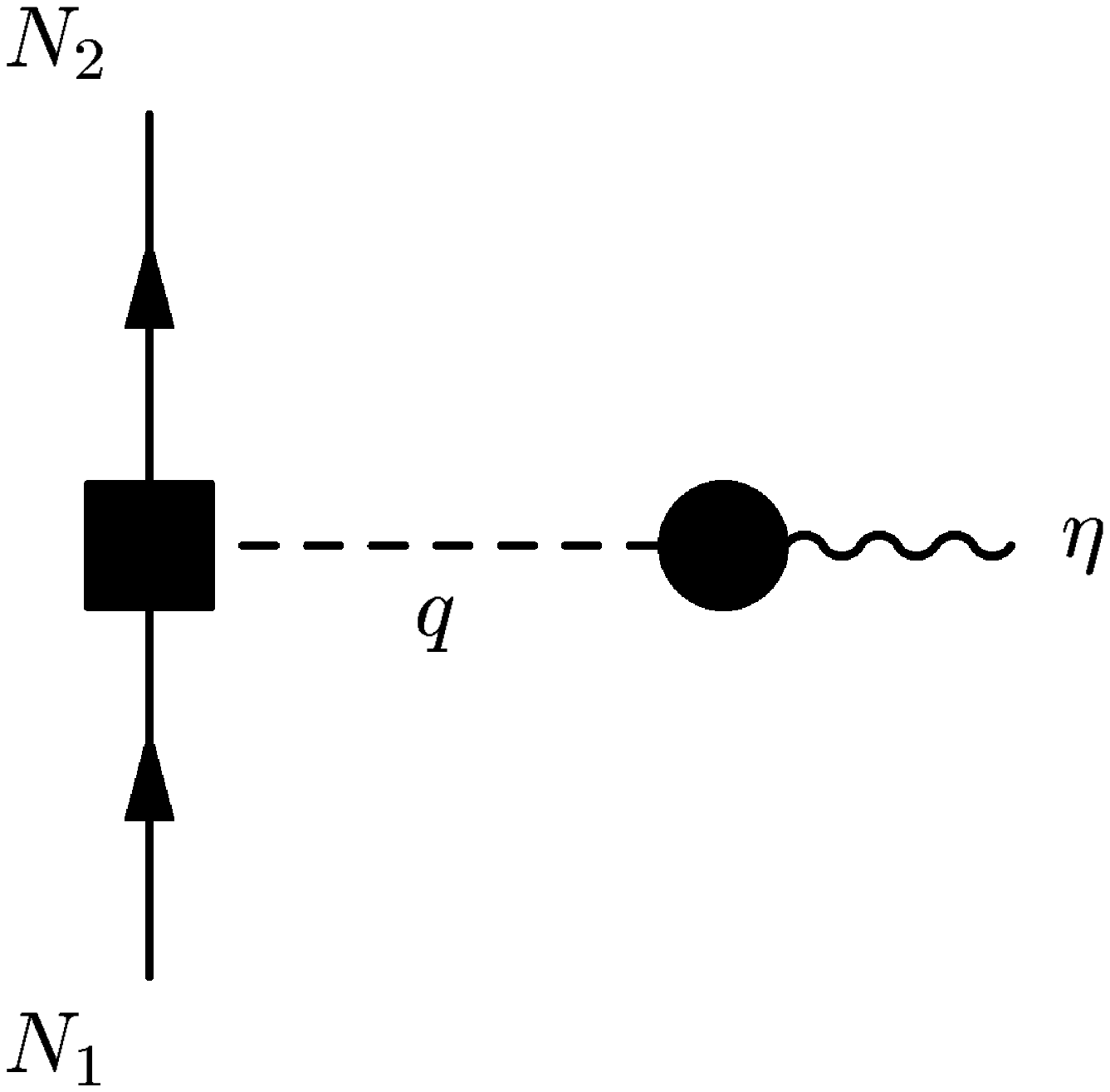,width=15cm}}
\vspace{0.1cm}
{\small Fig. 1}
\end{figure}
\begin{figure}[htbp]
\centerline{\psfig{figure=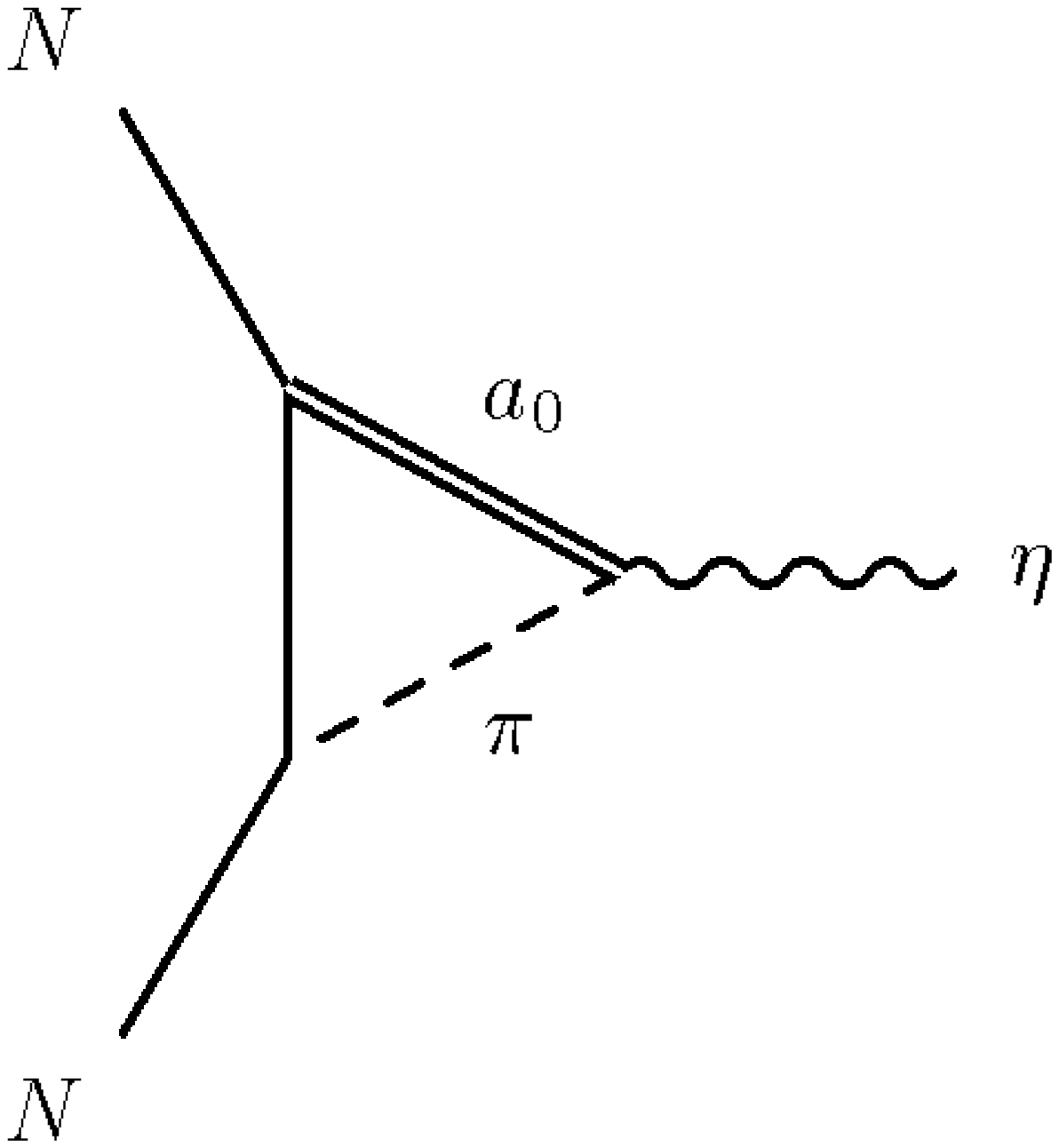,width=15cm}}
\vspace{0.1cm}
{\small Fig. 2}
\end{figure}
\begin{figure}[htbp]
\centerline{\psfig{figure=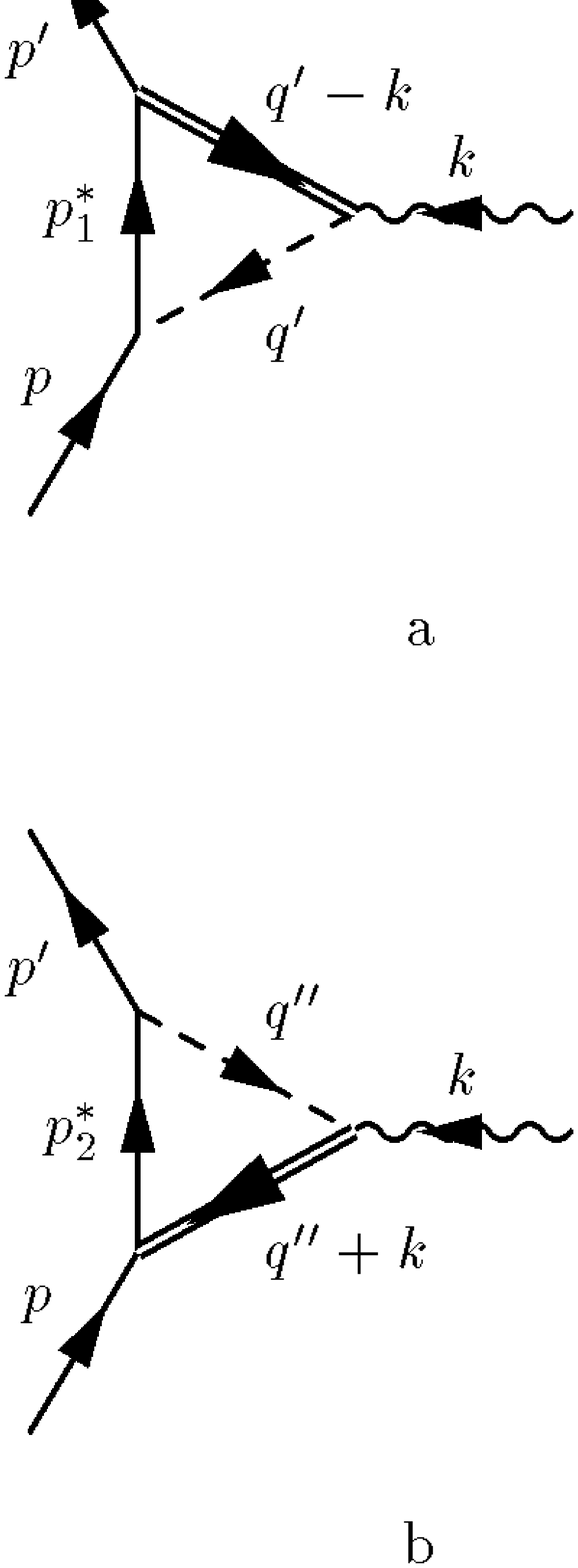,width=15cm}}
\vspace{0.1cm}
{\small Fig. 3}
\end{figure}
\begin{figure}[htbp]
\centerline{\psfig{figure=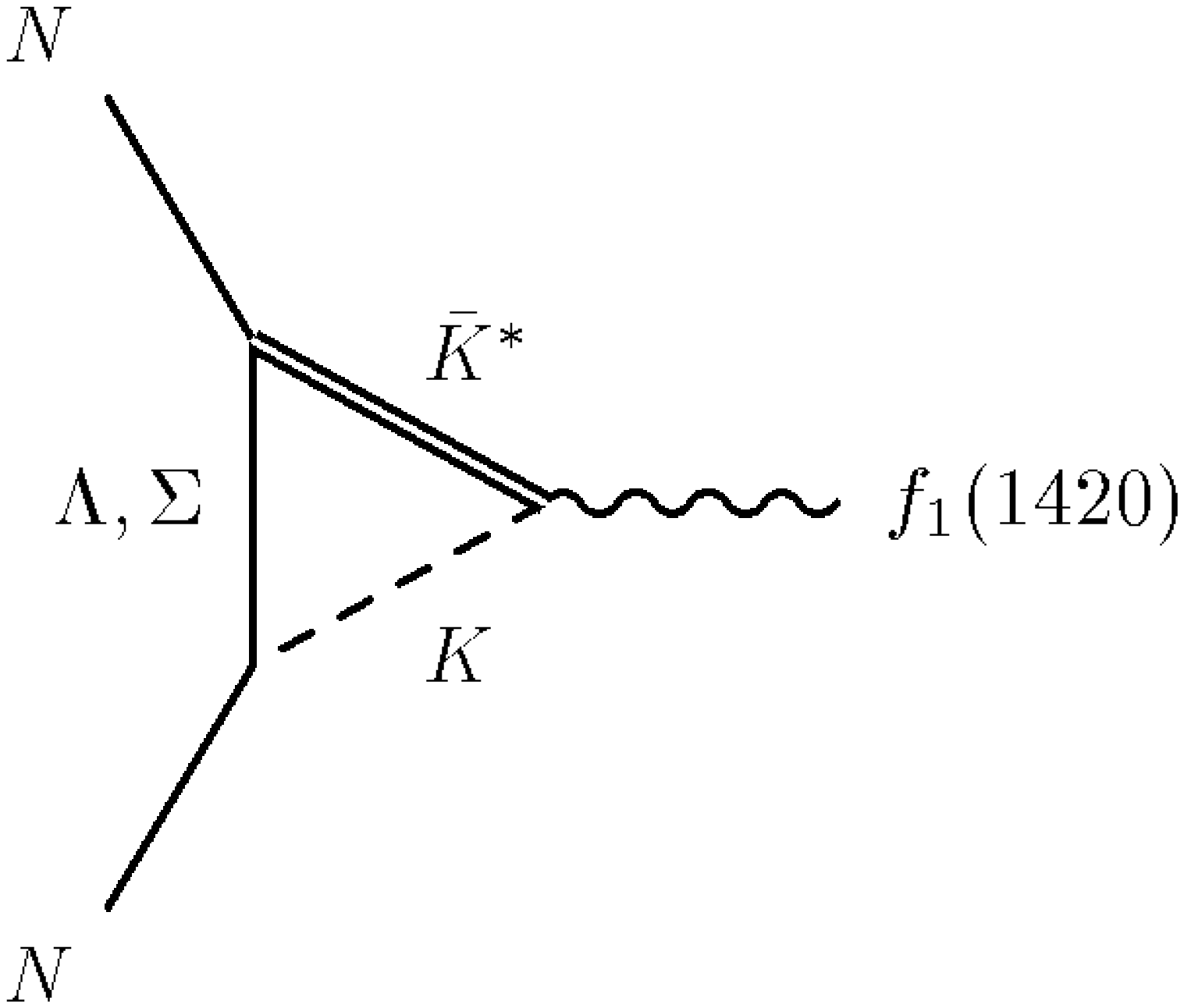,width=15cm}}
\vspace{0.1cm}
{\small Fig. 4}
\end{figure}
\begin{figure}[htbp]
\centerline{\psfig{figure=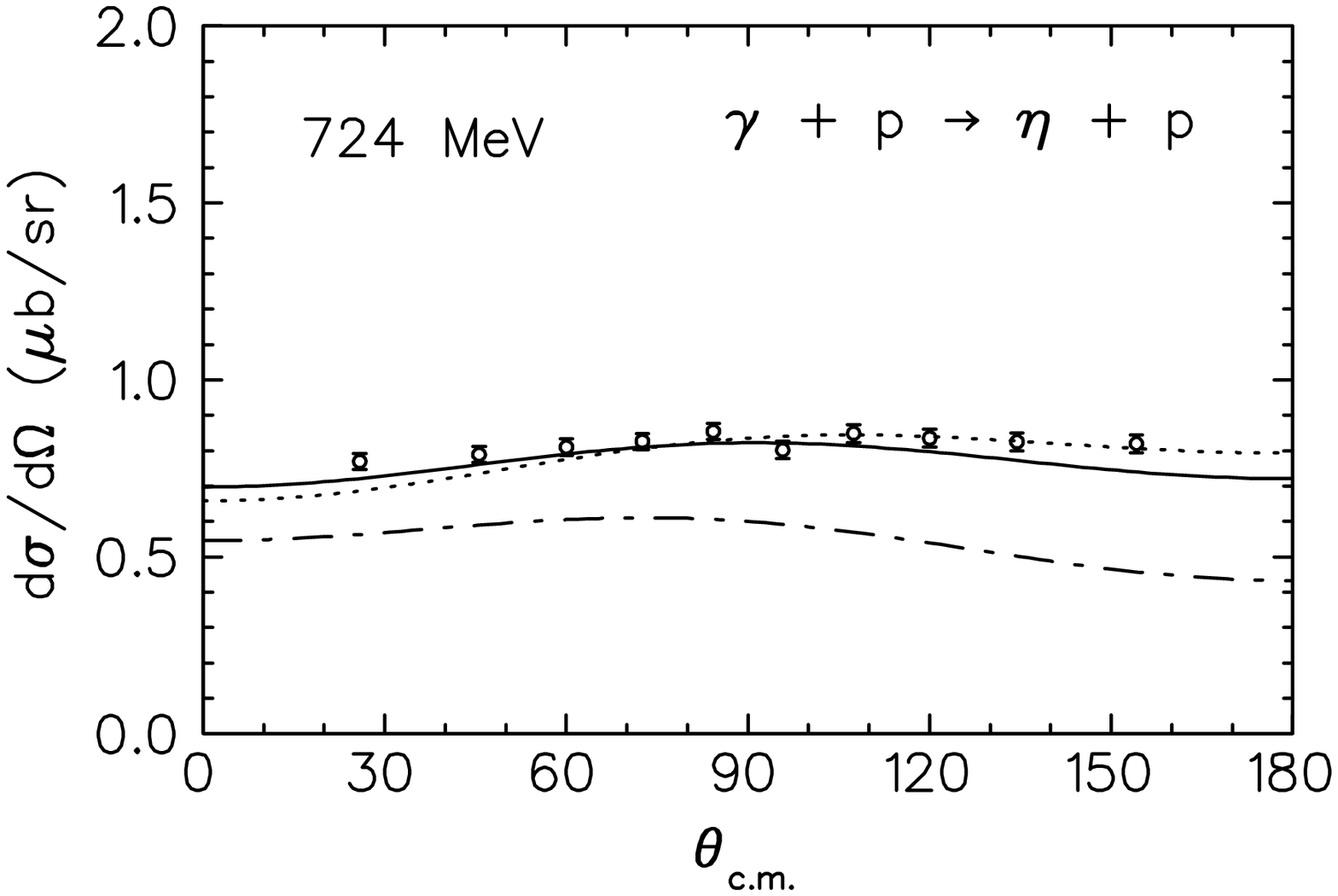,width=15cm}}
\vspace{0.1cm}
{\small Fig. 5}
\end{figure}

\end{document}